\documentclass{elsart3}

\usepackage{graphicx}
\usepackage{bm}

\newcommand{\be}{\begin{equation}}
\newcommand{\ee}{\end{equation}}

\newcommand{\bfr}{ {\bf r} }

\usepackage{amssymb}

\begin{document}

\begin{frontmatter}



\title{Path Integral Calculations of exchange in solid $^4$He}


\author{B. Bernu}
\address{LPTL, UMR 7600
of CNRS, Universit\'e P. et M. Curie, Paris, France}

\author{D. M. Ceperley}

\address{Dept. of Physics and NCSA, University of Illinois at
Urbana-Champaign, Urbana, IL 61801} \ead{ceperley@uiuc.edu}

\begin{abstract}
Recently there have been experimental indications that solid
$^4$He might be a supersolid. We discuss the relation of
supersolid behavior to ring exchange. The tunnelling frequencies
for ring exchanges in quantum solids are calculated using Path
Integral Monte Carlo by finding the free energy for making a path
that begins with the atoms in one configuration and ends with a
permutation of those positions. We find that the exchange
frequencies in solid $^4$He are described by a simple lattice
model which does not show supersolid behavior. Thus, the PIMC
calculations constrain the mechanism for the supersolid behavior.
We also look at the characteristics of very long exchanges needed
for macroscopic mass transport.
\end{abstract}

\end{frontmatter}


Recent torsional-oscillator observations by E. Kim and M. H. W.
Chan on solid $^4$He, both in the disordered absorbed vycor\ and
in bulk $^4$He \cite{chan}, have revived interest in the
super-solid phase.  A supersolid\cite{leggett} is characterized by
both long-range translational order and by a non-classical
response to rotation (NCRI). In attempting to understand these
results, one should first assume that the observed phenomena is
described by equilibrium thermodynamics of pure bulk solid helium.
For such a model, it is possible to perform numerically exact Path
Integral Monte Carlo simulations and explore its basic properties.
If such properties agree with experiments, that will provide
strong support for the model.  If one finds disagreement, then the
basic assumption is invalid. The results of such a calculation
have been recently published\cite{C181}. Here, we elaborate and
update that publication.

We consider the possibility that bulk solid hcp helium, assumed to
be free of defects, could have an equilibrium supersolid phase.
Because of its light mass, solid helium has large quantum zero
point fluctuations causing instantaneous ground state defects. At
any instant of time, a good fraction of atoms are closer to a
neighboring site than to their home site. However, the absence of
an atom from a lattice site is not sufficient for having a
supersolid;  if the empty site is always accompanied by doubly
occupied site, there can be no mass current. However,
Chester\cite{chester} proved that any Jastrow ({\it i.e.} pair
product) wavefunction of finite range has both BEC and vacancies.
Although Jastrow wavefunctions do not make a good description of
the ground state, one can multiply by a periodic trial
function\cite{C004,zhai} and achieve a crystalline, Bose condensed
state and a reasonable description of the energy, though not as
low as a wavefunction without BEC. Though this wavefunction of the
solid is not too bad for static properties, other properties,
particularly off-diagonal long range order (ODLRO), equivalent to
BEC, is not necessarily accurately given. There is no reason to
think that even a perfect match of the diagonal properties
constrains the off-diagonal properties.

Path integrals give a rigorous way of calculating on and
off-diagonal properties. The partition function of $N$ bosons is:
 \be  Z= \frac{1}{N!} \sum_P \int dR \langle R | e^{-\beta H} |
 PR \rangle \ee
where $H$ is the Hamiltonian, $\beta$ the inverse temperature and
$R= \{ \bfr_1, \bfr_2 , \ldots \bfr_N \}$. We will assume the
helium atoms interact with a semi-empirical\cite{aziz95} pair
potential. The helium potential is known to be accurate; many
calculations agree with experimental measurements, without the
adjustments needed in all other atomic and molecular systems.
Using the PIMC method, superfluidity and freezing, happen
naturally at the right density and temperature, without imposing
them in any way as is done in a wavefunction based approach.

To perform the PIMC calculations, the density matrix operator is
expanded into a path, $R(t)$ where t
($0\leq t \leq \beta$) 
is
imaginary time. Then the superfluid density ({\it i.e.} the number
of atoms not moving with the walls of the apparatus) is
proportional to the mean squared winding number, $\rho_s\propto
 <\vec{W}^2>$ where $\vec{W}=
\int_0^{\beta} dt \sum_{i=1}^N d\bfr_i(t)/dt $ is the winding
number of the paths  around the periodic boundary condition. Note
that only exchanges on the order of the sample size contribute to
the superfluid density; local exchanges make no contribution.
Prokof'ev and Svistunov\cite{prokofev} argue that long windings
are impossible without vacancies and interstitials that are
spatially separated.

The technical complications of path integral
calculations\cite{RMPI} concern ergodicity of the random walk, and
finite size effects: one has to take the limit as $N \rightarrow
\infty$. If one does a PIMC simulation of a 48 atom ($3 \times 4
\times 2$) hcp supercell in PBC, one finds a superfluid density of
about 3\% at melting density (molar volume 21.04 cm$^3$) and about
1.2\% at 55 bars (molar volume 19.01 cm$^3$). However, simulations
of a larger cell 180 atoms, show zero superfluid density. With
current algorithms, it is difficult to be sure that the lack of
winding paths has a physical origin, or is due to a lack of
ergodicity within the random walk. Winding number changes involve
the permutation of a group of particles which stretch across the
box. One needs to update simultaneously 5 atom coordinates to
change the winding number in the 180 cell. The acceptance
probability for such a collective move becomes very small. To
avoid the problem of estimating rare events, we turn to a PIMC
approach which directly estimates individual exchange
probabilities.

The Thouless\cite{thouless} theory of exchange in quantum crystals
assumes that at low temperatures, the system will almost always be
near one of the $N!$ arrangements of particles to lattice sites.
There will be rare, rapid, tunnellings from one arrangement to
another in an amount of ``imaginary time'' $\gamma_p$. Hence, we
can label the particles with their initial lattice sites and drop
the $N!$ in Eq. (1). The partition function is written as a sum
over permutations of lattice sites onto themselves.  We break up
the permutation into cyclic exchanges $P \equiv \{ p_1,p_2 \ldots
p_n\}$. We can then write:
 \be \label{ITP} Z= Z_0  \sum_P\prod_{i=1}^{n_P} f_{p_i} (\beta) . \ee
where $Z_0=\int dR \langle R | e^{-\beta H} | R \rangle $ is an
uninteresting phonon partition function at low temperature and the
contribution for an elementary cycle is:
 \be f_p (\beta) = \frac{1}{Z_0}\int dR \langle R | e^{-\beta H} | p R \rangle \equiv J_p \beta.
  \ee
The exchange is localized in imaginary time so the path integral
is  proportional to the number of crossing times: $f_p (\beta)
\propto \beta$ ; the coefficient, $J_p$ is the exchange frequency.
We have also assumed that independent cycles do not interfere with
each other, reasonable because the exchanges are rare and are
unlikely to be close in both space and imaginary time.

We have developed methods\cite{CJ,cornell} to calculate the
exchange frequencies in quantum crystals, primarily for solid
$^3$He. In that system, ring exchanges of 2-6 atoms give rise to
the magnetic properties at low temperatures. The PIMC calculations
were important in giving a microscopic justification of the
multiple exchange model. The PIMC values are in good agreement
with experimental data on $^3$He and are thus expected to be
accurate in the closely related solid $^4$He.

In Table 1 are shown the exchange frequencies for 2, 3 and 4 atom
exchanges in hcp $^4$He.  For these calculations, we use an
imaginary time step of $0.0125K^{-1}$ and a value of imaginary
time of $\beta=2K^{-1}$ (160 time slices).  For most of the
exchanges we find an exchange time $\gamma_p=0.3 K^{-1}$. We have
studied systems with 48 and 180 atoms; the larger value is
necessary for convergence of the exchange frequencies as is shown
in the table. The values are quite small {\it e. g.} $J_2\sim 3
\mu K $ at melting density.  We note that in an hcp crystal, there
are often several alternative ring exchanges of n
atoms\cite{roger}. For example, one can have a 2 particle
near-neighbor exchange in the basal plane $n$ or out of the basal
plane $n'$. Similarly a 3-body ring exchange can either enclose
atoms in nearby planes $T$ or not $T^*$. The frequencies for these
various alternatives can differ by an order of magnitude.

\begin{table}
\caption{Calculated exchange frequencies in hcp $^4$He at a molar
volume of 21.04 cc in units of $\mu K$. The notation of the
exchange is from Roger\cite{roger} as are the WKB actions for
$^3$He. The column $J_{48}$ are for $\beta=1$, $\tau=0.0125$ and
$N=48$ and $N=180$ for a larger cell. $\gamma_p$ is the exchange
time in units of $K^{-1}$.} \label{HE4J}\centering
\begin{tabular} { |c| c|c  c | c c|c|c |} \hline
p &  name & $J_{48}$&\% error& $J_{180}$ &\% error& $\gamma_p$&WKB \\
\hline
2 & nn    & 1.7 &  2 &3.74 &7&0.32  &9.54\\
  & nn'   & 2.5 &  6 &4.50 &5&0.31  &8.60\\ \hline
3 & T     & 5.6 &  4 &8.0  &5&0.29  &7.33\\
  & T*    & 0.62&  5 &1.25 &6&0.34  &8.83\\
  & T'    & 2.0 &  4 &4.9  &5&0.32  &7.75\\ \hline
4 & Kp'   & 0.74&  9 &0.59 &6&0.19  &9.00\\
  & Kp    & 0.84&  8 &0.71 &6&0.4   &9.38\\
  & Ksq'  & 1.25&  4 &2.2  &8&0.34  &8.09\\
\hline\end{tabular}
\end{table}

The small cyclic exchanges in Table 1 are quite different from the
long exchanges needed to get a supersolid. Accordingly, we have
performed exchange calculations of 50 different exchanges
involving from 5 to 10 atoms. All exchanges are between nearest
neighbors, since calculations show that next-nearest exchanges are
much less probable. We obtain accuracies on the order of 5\% for
the 5-particle exchanges and 10\% for the 9 particle exchanges.
More than half of the exchanges we studied involve winding around
the cell boundaries, important because they are representative of
the type of exchanges in a supersolid.

\begin{figure}
\includegraphics[width=7cm,height=7cm]{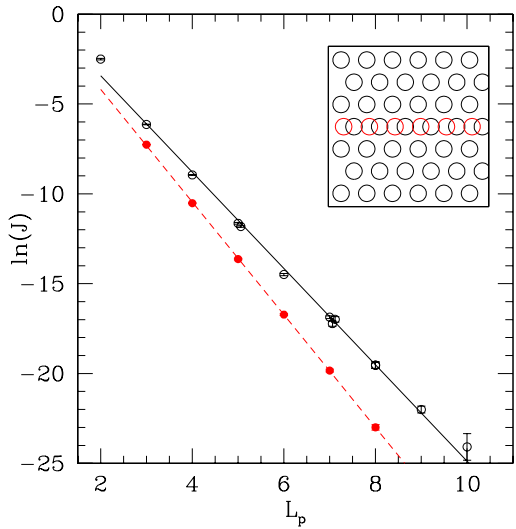}
\caption{\label{winding} The exchange frequencies (J in K) versus
exchange length $L_p$ for straight line exchanges in the basal
plane that wind around the periodic cell. The inset shows the
lattice sites in the basal plane for $L_p=6$; the red atoms show
the atoms midway through the exchange. The circles with error bars
are the $\ln(J)$ at two molar volumes; 21.04 cm$^3$ (black, open
circles, solid line) and 19.01 cm$^3$(red, solid circles, dashed
line). The lines are least squares fits.}
\end{figure}

Fig. \ref{winding} shows the results of calculations of the
frequency of the simplest winding exchanges: straight line
exchanges in the basal plane.  These calculations were done in an
hcp supercell by varying the number of unit cells in the
x-direction and keeping a fixed width in the y and z directions.
We find that the exchange frequencies decrease exponentially with
the length of the exchange with an exponent of $\alpha =2.64$ near
the melting volume 21.04 cm$^3$, and $\alpha=3.14$ at the volume
19.01 cm$^3$ corresponding to $P \sim$ 60 bars.

In order to understand whether the Bose-Einstein transition
occurred in superfluid $^4$He,  Feynman introduced a lattice model
\cite{feynman}.  The Feynman-Kikuchi (FK) partition function is:
\be  Z=  Z_0\sum_P  e^{ -\alpha \sum_{i=1}^{n_P} L_{p_i}} \ee
where the sum is over all closed non-intersecting polygons drawn
on a lattice by connecting nearest neighbors. Calculations were
done on a cubic 3D lattice, first by approximately by Feynman and
Kikuchi and then numerically by Elser\cite{elser}. It was found
that the superfluid transition occurs as $\alpha$ becomes smaller
from a localized state (small cycles) to large cycles for
$\alpha_c \approx 1.44$. To understand this critical value,
consider the free energy of adding a link to an existing loop.
Adding a link costs probability $e^{-\alpha}$ but the entropy of
the new link is $5$ in a cubic lattice; hence $\alpha_c \sim
\ln(5)=1.6$. The exact critical value is smaller because of the
'self-avoiding' restriction within a cycle and on overlapping
cycles. Using the same argument for the hcp lattice gives a
critical coupling of $\alpha_c \approx 2.3$. Our determined slope
in figure 1 is larger than this but not by much!

However, we need a more realistic model than the FK model to take
into account more details of the geometry of the exchange than
just the number of exchanging atoms; note for example the large
differences between the 2 types of nearest neighbor 4-atom
exchanges, a parallelogram $K_p$ and a square $K_{sq}$ in the
Table. We assume that it is only the internal geometry of the
exchange that matters\cite{lesh}; the detailed arrangement of the
neighboring spectator atoms is much less important. In particular,
we assume the the log of the exchange frequency is a sum over the
internal angles $\theta_k$ of the exchange:
 \be \label{model} J_p=J_0  \exp [-\sum_{k=1}^p\alpha(\theta_k)].\ee
In an hcp lattice there are 7 possible angles between two nearest
neighbor displacement vectors (neglecting possible dependence on
motion with respect to the basal plane), but $\theta=0$ only
occurs in the pair exchange which we neglect since it does not
contribute to the superfluid density. We determine the values of
the 7 parameters by fitting to the PIMC exchange frequencies
$J_p$. Good fits are obtained; the model predicts the exchanges
frequencies with an accuracy of about 20\%. (Note that the data
base does not include the small exchanges in table 1. The
frequencies for small exchanges are more sensitive to the
neighbors.) There is a strong preference for exchanges that
proceed in a straight line versus ones which have sharp angles;
for them the incoming and outgoing particles are more likely to
collide. We find that the fitting coefficients depend linearly on
$\cos^4(\theta/2)$ so that in the generalized F-K model we assume:
  \be J_p= J_o \exp[-\alpha
L_p-\alpha'\sum_{i=1}^{n_p} cos^4(\theta_i/2)]\label{GFK}.\ee

\begin{figure}
\includegraphics[width=7cm,height=7cm]{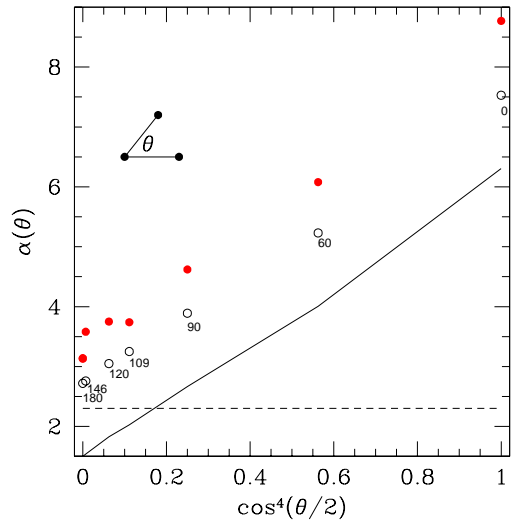}
\caption{\label{2pfitn}  The fitted angular dependance of the
action vs. $\cos^4(\theta/2)$ at the melting density , 25 bars
(open circles). The angle is the interior angle of a vertex as
shown. The fits were done using 50 different exchanges in a 180
atom cell. The solid red circles correspond to a pressure 60 bars.
The two lines show a critical supersolid Hamiltonian with (solid
line) and without (dashed line) angular dependance. }
\end{figure}

\begin{figure}
\includegraphics[bb= 100 350 600 700,width=7cm,height=4cm]{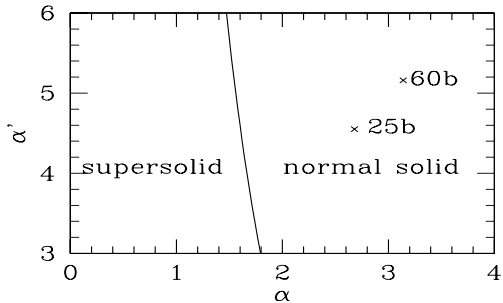}
\caption{\label{phased}  Phase diagram of the generalized
Feynman-Kikuchi model. The 2 points shown are for a density near
melting and at high pressure. }
\end{figure}

With the improved model, we can return to the question of whether
solid $^4$He is a supersolid. Because of the directionality of
exchange, the probability of retracing is small and we can ignore
the self intersections, allowing us to find the probability of
long exchanges by formulating it as a Markov process: the
diffusion of a particle in an hcp lattice. Note that on the basal
plane, exchanges can go straight ($\theta=180$), while an exchange
out of the basal plane must zig-zag ($\theta=146$), reducing the
exchange probability, thus the diffusion is anisotropic. Then, an
atomic displacement only depends on the previous atom's
displacement, not on the position in the lattice. (Because the hcp
lattice has a basis, we have to label the displacements
consistently on the A and B planes so that the transition
probabilities are independent of the plane.)
Let the (un-normalized)
probability of a given displacement vector after $n$ steps be
$\Pi_i^{(n)}$ where $i$ refers to one of the lattice directions.
Then it will satisfy the master equation
 \be \Pi_j^{(n+1)}= \sum_{i} \Pi_i^{(n)} e^{\alpha(\theta_i-\theta_j)}\ee
The probability for long exchanges will correspond to the maximum
eigenvalue $\lambda$ of its stationary state: \be \sum_{i} \Pi_i
e^{\alpha(\theta_i-\theta_j)}=\lambda \Pi_j. \ee By symmetry
$\Pi_i$ can only depend on whether the direction is in the basal
plane, $\Pi_1$, or out of the basal plane, $\Pi_2$. Define the
partial sum $D_{i,j}=\sum_{j}e^{-\alpha(\theta_i-\theta_j)}$ where
$(i,j)$ are either in the basal plane (1) or out of it (2), {\it
e.g.} $D_{11}$ is the sum of the probability of both vectors being
in the basal plane). The eigenvalue $\lambda$, is the solution to
the secular equation:
 \be ( D_{11}-\lambda)(D_{22}-\lambda)-D_{12}^2=0. \ee
\be    \lambda =(D_{11}+D_{22})/2+D_{12}[1+y^2]^{1/2} \ee where
$y=(D_{22}-D_{11})/(2D_{12})\approx-0.023.$ Putting in the model
parameters, we find $\lambda= 0.303 \pm 0.005$. The probability of
having an exchange of length $n$ will equal a prefactor times
$\lambda^n$, hence,  since $\lambda<1$, our PIMC results imply
that solid $^4$He  will have only localized exchange and thus
cannot be a supersolid. Since $y$ is small, the diffusion in and
out of the basal plane are similar. Self-intersections and the
presence of other exchange cycles will further decrease the
probability of long exchange cycles and increase the critical
value of $\lambda$. (To estimate the effect of self-intersections
quantitatively, we performed random walks on the hcp lattice and
counted non-intersecting walks.  This line is shown in Fig.
\ref{2pfitn}.) As seen on Fig. \ref{phased} the estimated value of
$\lambda$ is much less than what is needed to allow for a
supersolid. In addition, there is no indication that long
exchanges become prevalent near melting.

\begin{table} \caption{Parameters for the generalized F-K exchange model, Eq. \ref{model}  as
determined by fitting to PIMC determined exchange frequencies.}
\centering
\begin{tabular}{|c|c|c|c|c|c|}
  \hline
 v (cm$^3$) & P (bar)& $J_0$(K )& $\alpha$ & $\alpha'$ & $\lambda$ \\
  \hline
  21.04 & 25 &7.2 & 2.64& 4.8 & 0.31 \\
  19.01 & 60 &8.0 & 3.14 & 4.8 & 0.25 \\
  \hline
\end{tabular}
\end{table}

One assumption made in the calculations is that long exchanges are
correctly being treated by the method to calculate $J$. We find
that the exchange time is roughly independent of the number of
exchanging atoms, implying that an exchange of $L_p$ atoms happens
by all of the atoms simultaneously sliding over to their new
positions. As $L_p$ increases, such a mechanism seems unlikely.
There would be much more phase space for the exchange path if
different portions of a long path exchanged at different times.

To understand the mechanism for long exchanges, we have done
special simulations of them. Though it is not possible to
calculate the frequency by the existing algorithm, it is
straightforward to sample long exchange paths. In particular, we
have studied the exchanges shown in fig. \ref{winding}, where
$L_p$ atoms slide along the basal plane in the x-direction, where
$10 \leq L_p \leq 50$. The boundary conditions in imaginary time
are $R(0)=Z$ and $R(\beta)=PZ$ where $P$ is a cyclic exchange of
atoms along a straight line. (The cell has a fixed extent in the y
and z-direction.) We do single particle updates on all the atoms
using the third level bisection procedure. Because of the ergodic
problem, the initial exchange will never decay, so we can study it
at will. In particular, we are interested in whether a long
exchange maintains the instanton character referred to above, or
whether it becomes delocalized in time.

\begin{figure}
\includegraphics[bb= 10 30 370 400,width=7.5cm,height=8cm,clip=true]{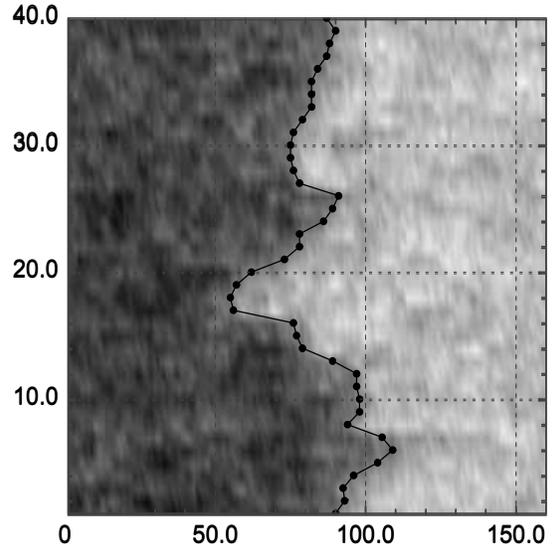}
\caption{\label{reaction}Gray scale image of the reaction
coordinate of a winding path containing 40 atoms, as a function of
particle number (y axis) and imaginary time (x-axis). The darker
region on the left side have coordinates closer to the initial
lattice sites, the lighter region on the right side is closer to
the permuted lattice. The black dots show the reaction time: when
each atom is midway between its initial and final positions. A
vertical portion of the reaction time curve (such as at the top)
represents a group of atoms that move as a unit. Horizontal (or
jagged) parts of the curve represent interstitials. }
\end{figure}

In order to quantify the exchange process, we define a reaction
coordinate of each atom involved in the exchange as $\eta_i(t)
=-1/2+ (x_i (t)-x_i(0))/a$ where $a$ is the nearest neighbor
distance and $x_i(t)$ is the $x$-coordinate of atom $i$ at
imaginary time $t$. The reaction coordinate is defined so that
$\eta_i(0)=-1/2$ and $\eta_i(\beta)=1/2$. Fig. \ref{reaction} is a
gray scale image of the reaction coordinates for a typical path
with $L_p=40$. One sees a fairly sharp division between the
initial lattice sites (in black on the left) and the final lattice
position (light color on the right). We define the exchange time
$e_i$ for a given atom as the solution to $\eta_i (e_i) = 0$. In
case of multiple solutions we find the solution where the
neighbors atoms are also exchanging. The exchange times for the
path in fig. \ref{reaction}  are shown as dots and connected with
a solid line.

The resulting curve of exchange times can be used to characterize
the exchange process. Nearly vertical portions, represent
instantons. In the figure,  16 atoms (labelled on the y axis as
27-40 1-3 
) are such a segment and the 5 atoms (8-12) another. However, in
long exchanges, such segments are broken by an interstitial
vacancy (i-v) formation, a subsequent pinning of the interstitial
and diffusion of the vacancy. An example is atom 16 followed by
the diffusion of the vacancy (atoms 17-26). The exchange rarely
proceeds by a single i-v event. The jagged appearance of the
exchange curve is confirmation that vacancies and interstitials
are strongly bound and cannot be separated by more than a few
lattice spacings.  This is different than the creation of a single
i-v pair that is discussed in Prokof'ev and B.
Svistunov\cite{prokofev}. In the path represented in fig.
\ref{reaction}, at least 3 i-v fluctuations  and 4 or more
instantons are present. By forcing the long exchange, we create
multiple i-v pairs as in a recently proposed supersolid
mechanism\cite{ma}. However, such i-v pairs are not low energy
excitations; otherwise they would have been detected in specific
heat or other thermodynamic measurements\cite{meisel}.

\begin{figure}
\includegraphics[bb= 80 450 500 718,width=6cm,height=3.5cm]{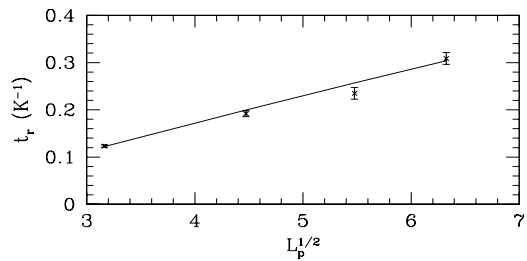}
\caption{\label{trough} The ''roughness'' of the exchange time
curve versus the number of exchanging atoms. See the text for the
definition of roughness. The solid curve is a calculation of a
simple Brownian bridge model of the roughness.  }
\end{figure}

\begin{figure}
\includegraphics[bb= 80 450 500 718,width=6cm,height=3.5cm]{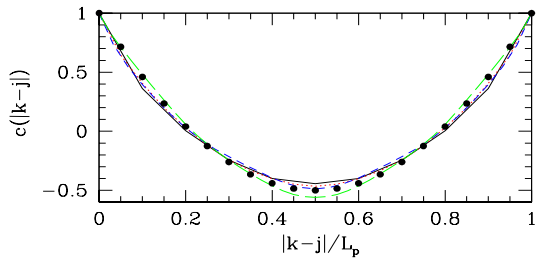}
\caption{\label{correlation} The auto-correlation (see text) of
the exchange time vs separation of atoms. Shown are the results
for $L_p$=10 (black), 20(red), 30(blue) and 40(green). The curve
goes negative for large cycles because it must average to zero and
has periodicity $L_p$. The dots are analytic results for a
Brownian bridge.}
\end{figure}

To quantify this process, we examine the ``roughness'' of the
exchange curve versus the system size.  We characterize the
roughness by first sorting the exchange times and discarding the
largest and smallest $L_p/10$ values to make the statistics more
robust. The roughness is then $t_r = \left< e_{\max}-e_{\min}
\right>$. The roughness grows as $L_p^{1/2}$, see fig.
\ref{trough}. Also shown is the same measure of roughness computed
on a Brownian bridge with $L_p$ elements (random 1D diffusion
returning to the start after $L_p$ steps). Figure
\ref{correlation} shows the autocorrelation function $c(k-j)=
\left< \delta e_k \delta e_j \right>/t_r^2$ of the exchange time
for systems with $10 \leq L_p \leq 40$. The autocorrelation
function is very close to that given by that of a Brownian bridge:
$c(x)=6x^2-6x+1$. This is another indication that the exchange
time is a random diffusion process. Hence, for long exchanges, the
exchange time should grow as $L_p^{1/2}$, because of the creation
of a i-v pairs every 5 to 10 atoms. The decay rate of the exchange
frequency will still be exponential: $J= \exp(-\alpha L_p)$ since
such a law results from both an instanton process and the creation
of multiple i-v pairs.

In summary, PIMC-computed exchange frequencies for hcp solid
$^4$He show that only localized exchanges will be present and thus
should not exhibit the property of nonclassical rotational
inertia. We find no indication that long cycles behave differently
than short cycles. In a separate study, we modelled $^4$He
absorbed in Vycor\cite{C185} and found that a persistent liquid
layer picture could explain the experimental findings.  To explain
the bulk $^4$He experiment one must look for other mechanisms,
either non-equilibrium dynamical effects or more complicated
lattice defects not allowed by the boundary conditions in our
simulations.

This work was supported by CNRS and the fundamental physics
program at NASA (NAG-8-1760). We also thank V. Blancheton and E.
Masahiro for running the code that provided data for figs. 3-4.

\end{document}